\documentclass{pnastwo}

\usepackage{graphicx}

\usepackage{pnastwof}

\usepackage{amssymb,amsfonts,amsmath}
\usepackage{times}

\contributor{submitted}
\url{www}
\copyrightyear{2008}
\issuedate{Issue Date}
\volume{Volume}
\issuenumber{Issue Number}

\begin{document}


\title{Inferring fitness landscapes by regression produces biased estimates of epistasis}





\author{Jakub Otwinowski\affil{1}{Department of Biology, University of Pennsylvania, Philadelphia PA, 19104} Joshua B. Plotkin\affil{1}{}}

\contributor{Submitted to Proceedings of the National Academy of Sciences
of the United States of America}

\maketitle

\begin{article}

\begin{abstract} The genotype-fitness map plays a fundamental role in shaping the
dynamics of evolution. However, it is difficult to directly measure a fitness
landscape in practice, because the number of possible genotypes is astronomical.
One approach is to sample as many genotypes as possible, measure their fitnesses,
and fit a statistical model of the landscape that includes additive and pairwise
interactive effects between loci. Here we elucidate the pitfalls of using such
regressions, by studying artificial but mathematically convenient fitness
landscapes. We identify two sources of bias inherent in these regression
procedures that each tends to under-estimate high fitnesses and over-estimate low
fitnesses. We characterize these biases for random sampling of genotypes, as well as
for samples drawn from a population under selection in the Wright-Fisher model of
evolutionary dynamics. We show that common measures of epistasis, such as the
number of monotonically increasing paths between ancestral and derived genotypes,
the prevalence of sign epistasis, and the number of local fitness maxima, are
distorted in the inferred landscape. As a result, the inferred landscape will
provide systematically biased predictions for the dynamics of adaptation.  We
identify the same biases in a computational RNA-folding landscape, as well as in
regulatory sequence binding data, treated with the same fitting procedure.
Finally, we present a method that may ameliorate these biases in some cases.
\end{abstract}

\keywords{term | term | term}





\section{Significance statement} The dynamics of evolution depend an organism's
``fitness landscape", the mapping from genotypes to reproductive capacity.
Knowledge of the fitness landscape can help resolve questions such as how
quickly a pathogen will acquire drug resistance, or by what pattern of mutations.  
But direct measurement of a
fitness landscape is impossible because of the vast number of genotypes.
Here we critically examine regression techniques used to approximate fitness
landscapes from data.  We find that such regressions are subject to two inherent
biases that distort the biological quantities of greatest interest, often making
evolution appear less predictable than it actually is.  We discuss methods that
may mitigate these biases in some cases.

\section{Introduction}

An organism's fitness, or expected reproductive output, is determined by its
genotype, environment, and possibly the frequencies of other
genotypes in the population. In the simplified setting of a fixed environment, and
disregarding frequency-dependent effects, which is typical in many
experimental populations \cite{Lenski1991, Lenski1994,
Elena2003,Blount2008,Woods2011}, fitnesses are described by a map from genotypes
to reproductive rates, called the \emph{fitness landscape}.

The dynamics of an adapting population fundamentally depend on characteristics of
the organism's fitness landscape
\cite{Kingman,Kauffman1987,Kauffman1989,Macken1989,Flyvbjer1992,Perelson1995,Newman1998,Orr2005,Cowperthwaite2007,Park2008,Phillips2008,Tokuriki2009,
Poelwijk2007,Kryazhimskiy2009a,Bridgham2009,Bloom2009,Lunzer2010,Martinez2011,Novais2010}.
However, mapping out an organism's fitness landscape is virtually impossible in
practice because of the coarse resolution of fitness measurements, and because of
\emph{epistasis}: the fitness contribution of one locus may depend on the states
of other loci. To account for all possible forms of epistasis, a fitness landscape
must assign a potentially different fitness to each genotype, and the number of genotypes
increases exponentially with the number of loci. 

As a result of these practical difficulties, fitness landscapes have been directly
measured in only very limited cases, such as for individual proteins, RNA
molecules, or viruses.  Even in these limited cases genetic variation was restricted to a
handful of genetic sites
\cite{Burch1999,Lee1993,Lee1997,Schlosser2005,Zhang2009,Hayden2012,Weinreich2006,Reetz2008,Lozovsky2009,Hietpas2011,Hall2010,Chou2011,Khan2011a,Remold2004,Rokyta2011,Trindade2009,Kvitek2011a}.
Alternatively, one might try to infer properties of a fitness landscape from a
time-series of samples from a reproducing population. Despite considerable effort
along these lines \cite{Kryazhimskiy2009a, Illingworth2012, Kryazhimskiy2011a,Illingworth2011},
this approach is difficult and such inferences from times-series can be subject to
systematic biases \cite{Draghi2013}.  As a result, very little is known about fitness
landscapes in nature, despite their overwhelming
importance in shaping the course of evolution. 

Technological developments now allow researchers to assay growth rates of microbes
or enzymatic activities of individual proteins and RNAs for millions of variants
\cite{Pitt2010,Jacquier2013,Jimenez2013}. As a result, researchers are now
beginning to sample and measure larger portions of the fitness landscapes than
previously possible. Nonetheless, even in these cases, the set of sampled
genotypes still represents a tiny proportion of all genotypes, and likely
also a tiny proportion of all viable genotypes. 

In order to draw conclusions from the limited number of genotypes whose fitnesses
can be assayed, researchers fit statistical models, notably by penalized regression,
that approximate the fitness landscape based on the data available.
This situation is perhaps best illustrated by recent studies of fitness for the
HIV-1 virus, based on the measured reproductive capacity of HIV-derived amplicons
inserted into a resistance test vector \cite{Hinkley2011, Kouyos2012}. These HIV
genotypes were sampled from infected patients. (An alternative approach, often
used for measuring activities of an individual enzyme, is to introduce mutations
randomly into a wild-type sequence
\cite{Fowler2010,Kinney2010a,Araya2011,McLaughlinJr2012,Jacquier2013}).
Whereas the entire fitness landscape of HIV-1 consists of reproductive values for
roughly $2^{1800} \approx 10^{600}$ genotypes, only $\approx 70,000$ genotypes
were assayed in the experiment \cite{Hinkley2011}. Researchers therefore
approximated the fitness landscape by penalized regression, based on the
measured data, using an expansion in terms of main effects of loci and epistatic
interactions between loci.  The principal goal of estimating the underlying fitness
landscape was to assess the extent and form of epistasis \cite{Hinkley2011}, and,
more generally, to understand how adaptation would proceed on such a landscape
\cite{Kouyos2012}. 

These \cite{Hinkley2011, Kouyos2012} and other high-throughput fitness
measurement studies \cite{Pitt2010,Jacquier2013,Jimenez2013} produce
massive amounts of data, but not nearly enough to determine an entire
fitness landscape.  This presents the field with several pressing questions: Do
statistical approximations based on available data faithfully reproduce the
relevant aspects of the true fitness landscape and accurately predict the dynamics
of adaptation? Or, do biases arising from statistical fits or measurement noise
influence the conclusions we draw from such data?


Here, we begin to address these fundamental questions about empirical fitness
measurements and how they inform our understanding of the underlying fitness
landscape and evolution on the landscape.
We study the effects of approximating a fitness landscape from data in terms of
main and epistatic effects of loci.  We demonstrate that such approximations,
which are required to draw any general conclusions from a limited sample of
genotypes, are subject to two distinct sources of biases. Although these biases
are known features of linear regressions, they have important consequences for the
biological quantities inferred from such fitness landscapes. These biases
systematically alter the form of epistasis in the inferred fitness
landscape compared to the true underlying landscape. In particular, the inferred
fitness landscape will typically exhibit less local ruggedness than the true landscape,
and it will suggest that evolutionary trajectories are less predictable than they
actually are in the true landscape. 

Most of our analysis is based on samples from mathematically constructed fitness
landscapes. But we argue that the types of biases we identify apply generally, and
in more biologically realistic situations. Indeed, we show that the same
types of biases occur in RNA-folding landscapes as well as empirically measured
regulatory sequence binding landscapes.

Although it may be impossible to completely remove these biases, we conclude by
suggesting steps to mitigate the biases in some cases.

\section{Results}
\subsection{Statistical approximations of fitness landscapes}

A function that maps genotype to fitness may be written as an expansion in terms
of main effects and interactions between loci \cite{Stadler1999,Hansen2001,Neher2011,Neidhart2013a,Weinreich2013}: 
\begin{equation}
y=\beta_{0}+\sum_{i}^{L}\beta_{i}x_{i}+\sum_{i<j}\beta_{ij}x_{i}x_{j}+\ldots,\label{eq:quad}
\end{equation} 
where $x_{i}$ represents a genetic variant at locus $i$, $y$ is the
fitness (typically the logarithm of growth rate), and $L$ is the number
of loci.  The nucleotides ATGC, or any number of categorical variables, may be
encoded by dummy variables, represented by $x_{i}$, which equal either
$+1$ or $-1$ in this study (see Methods). The term
$\sum_{i<j}$ represents a sum over all pairs of interactions between loci, and the
elipses represent higher-order terms, such as three-way interactions.
Since the statistical model is linear in the coefficients $\beta_i$ and
$\beta_{ij}$, etc, the best-fit coefficients can be inferred by linear regression.

Experimental data are now sufficiently extensive that both the additive and
pairwise epistatic coefficients, $\beta_i$ and $\beta_{ij}$, can often be estimated,
whereas three-way and higher-order interactions are typically omitted from the
statistical model of the fitness landscape. We refer to a statistical model with only
additive and pairwise interactions as a quadratic model.  Even in the quadratic
case, the statistical model may involve more free coefficients than empirical
observations, so that over-fitting could become a problem.  Techniques to accommodate
this problem, and the biases they introduce, are discussed below.

\subsection{Bias arising from penalized linear regressions}

The first type of bias we study arises from the use of penalized regressions --
which are required when a large number of parameters must be inferred from a
limited amount of data. Under standard linear regression with limited data,
overfitting can cause the magnitudes of inferred coefficients to be large,
resulting in positive and negative effects that cancel out to fit the observed
fitness measurements. The standard remedy for overfitting is a so-called
``penalized least-squares regression", such as ridge  or LASSO
regression \cite{Friedman2001}, which constrains the complexity of the inferred model by limiting the magnitudes of the inferred coefficients. For
example, in fitting a quadratic landscape to sampled HIV-1 fitness measurements,
Hinkley et al. employed a form of penalized linear regression in order to avoid
overfitting their data \cite{Hinkley2011}.


Although often required when fitting complex fitness landscapes to data, the
penalized least square regression has some drawbacks.  In general, the mean square
error of any regression can be decomposed into a bias and a variance. The
Gauss-Markov theorem guarantees that the standard least-squares linear regression
produces the lowest possible mean squared error (MSE) that has no bias, whereas
penalized least squares can reduce the MSE further by adding bias in exchange for
a reduction in variance \cite{Friedman2001}.  Thus, in order to provide
predictive power for the fitnesses of un-observed genotypes, these regressions
necessarily produce biased fits. While the accuracy of predicting unobserved
fitnesses may be improved by such a biased fit, other quantities of biological
interest derived from these predictions, such as measures of epistasis, may be
distorted by the bias.

In order to quantify the biases introduced by penalized least square regression,
we compared mathematically constructed fitness landscapes to the landscapes
inferred from a quadratic model fit by ridge regression (similar results hold for
LASSO regressions, see Discussion). Our analyses are based on two types of
mathematical fitness landscapes. The widely used $NK$-landscapes of Kauffman et al
\cite{Kauffman1987,Kauffman1989,Macken1989,Flyvbjer1992} comprise a family of
landscapes that range from additive to highly epistatic, depending upon the
parameter $K$, which determines the number of (typically sparse) interactions
between sites. We also study ``polynomial" landscapes, which consist of additive
effects and all possible pairwise and three-way interactions. In these landscapes,
the amount of epistasis can be tuned by controlling the proportion of fitness
variation that arises from the additive contributions, pairwise interactions, and
three-way interactions (see Methods). 

\begin{figure}
\begin{centering}
\includegraphics[width=1\columnwidth]{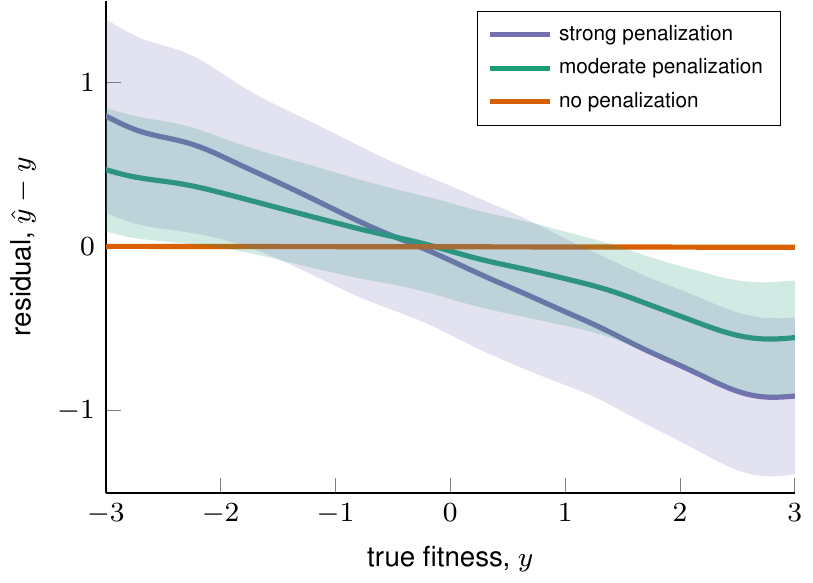}
\par\end{centering}
\caption{Fitness-dependent bias due to penalized regression. 
Penalized regression tends to reduce the magnitude of inferred coefficients, which
biases the estimated fitness, $\hat{y}$, towards the average value, so that the
high fitnesses are underestimated and low fitnesses are overestimated.
The plot shows the mean (solid line) and standard
deviation (shaded area), of the distribution of residuals at a given true fitness
value $y$, smoothed
by a gaussian moving window (see Methods). The fewer the observations (i.e.~the
smaller the number of genotypes sampled for fitting the statistical model), the
stronger the effect of this bias, as seen by comparing fits to training datasets
of different sizes: 250 sampled genotypes (red), 200 sampled genotypes (green),
and 150 sampled genotypes (blue). Genotypes were sampled randomly from a
quadratic polynomial fitness landscape, which lacks any three-way interactions
(parameters $v_1=2/3$, $v_2=1/3$, $v_3=0$, $\sigma^2_y=1$, and $L=20$ sites; see Methods). The
training data were fit to a quadratic model, which has $p=211$ parameters, 
so that the statistical model is well-specified. A test set of 5,000 random
genotypes was used to compare the predicted ($\hat{y}$) and true ($y$) fitnesses
of genotypes. With sufficient sampled data, no penalization is required and the
resulting statistical fit contains no bias (red).  \label{fig:regbias}}
\end{figure}

We constructed $NK$ and polynomial landscapes with only additive and pairwise effects, we sampled genotypes and
fitness from these landscapes, and we fit a quadratic model of the landscape based
on the sampled ``data" using penalized least square regression. For both $NK$ and
polynomial landscapes, we found that the inferred landscape tends to overestimate the fitnesses of
low-fitness genotypes and underestimate the fitnesses of high-fitness genotypes
(polynomial landscape Fig.~\ref{fig:regbias}, $NK$ landscape
Fig.~\ref{fig:regbias_NK}).  Thus, there is a fitness-dependent bias in the
inferred landscape compared to the true underlying landscape. The extent of
this bias depends on the amount of penalization used, which in turn depends on the
amount of data sampled relative the number of free coefficients in the statistical
model. When the number of independent samples equals the number of free
coefficients these biases disappear (Fig.~\ref{fig:regbias}, red curve), but
whenever data are in short supply these biases arise and they can be substantial
in magnitude (Fig.~\ref{fig:regbias}).

The only way to avoid this bias entirely is to obtain at least as many independent
observations as model parameters, which is typically unfeasible for realistic
protein lengths or genome sizes. 
Furthermore, as we will discuss below, these inherent biases have important
consequences for our understanding of epistasis in the fitness landscape and for
our ability to predict the dynamics of adaptation.

\subsection{Bias arising from a mis-specified model}

Even when there is sufficient data so that a penalized regression is unnecessary,
there is another source of potential bias in the inferred fitness landscape due to
variables that are omitted from the statistical model but present in the true
landscape, e.g.~higher-order interactions between loci \cite{Weinreich2013}.  In this case, the estimated coefficients of the statistical
model will be biased in proportion to the amount of correlation between the
omitted variables and the included variables \cite{Wooldridge2009}. Uncorrelated
omitted variables, by contrast, may be regarded as noise, as we discuss below.

Interactions of different orders, e.g.~three-way and pairwise interactions, are
generally correlated with each other, unless the genotypes are sampled randomly
and encoded as $\pm 1$ forming an orthogonal basis
\cite{Weinberger1991,Neher2011}.  In this case, which rarely applies to samples
drawn from an evolving population, the omitted interactions may be regarded as
noise and the estimated coefficients are guaranteed to be unbiased. However, even
in this case the inferred $y$ values may still be biased. 

\begin{figure}
\begin{centering}
\includegraphics[width=1\columnwidth]{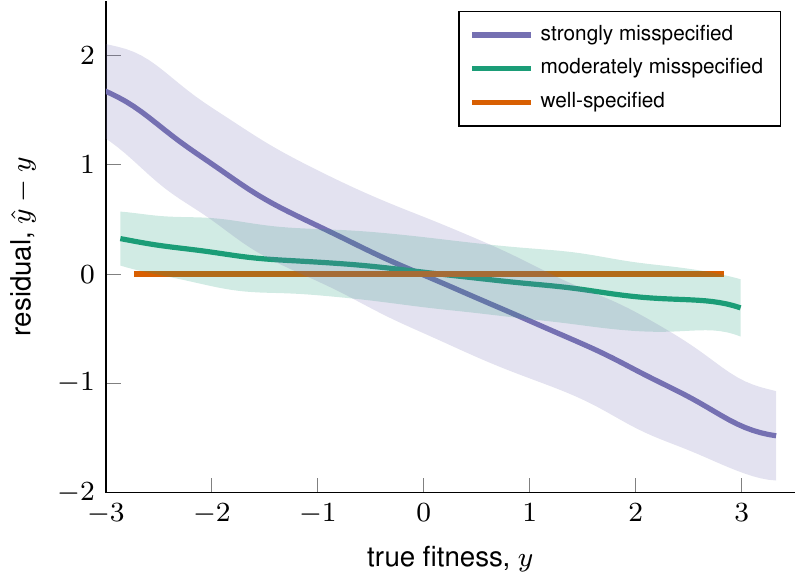}
\par\end{centering}
\caption{Fitness-dependent bias due to model misspecification. 
A mis-specified statistical model of the fitness landscape tends to bias 
predicted fitnesses towards the mean fitness,
resulting in under-estimated high-fitness genotypes, and
over-estimated low-fitness genotypes.
The figure is based
on quadratic fits ($p=211$ parameters) to 5,000 randomly sampled individuals 
from three different cubic-polynomial landscapes each with 
with $L=20$ sites and $\sigma^2_y=1$: $v_1=1/3$, $v_2=1/3$, $v_3=0$ (red), $v_1=0.6$, $v_2=0.3$,
$v_3=0.1$ (green), and $v_1=1/3$, $v_2=1/6$, $v_3=1/2$ (blue). The larger the
value of $v_3$, the greater the amount of model mis-specification and the stronger
the bias. 
A test set of 5,000 random genotypes was used to compare the predicted 
($\hat{y}$) and true ($y$) fitnesses of individuals.  \label{fig:modelbias}
}
\end{figure}

Fig.~\ref{fig:modelbias} illustrates the biases arising from model
mis-specification. To produce this figure we fit quadratic models to fitness
landscapes that contain higher-order interactions. In both the cubic polynomial
(Fig.~\ref{fig:modelbias}) and $NK$ (Fig.~\ref{fig:modelbias_NK}) landscapes,
fitnesses that are very high or very low are likely to contain contributions from
higher-order interactions with positive or negative effects, respectively. But
these higher-order interactions are not estimated by the statistical model, and so
the inferred model overestimates low fitnesses and underestimates high fitnesses.
Bias arising from model mis-specification is qualitatively similar to bias
arising form penalized regression, discussed above.  Bias from a mis-specified model
can be large, but it would be not be visible in a plot of residuals versus
\emph{inferred} fitnesses. 

The mis-specified model bias shown in Figure \ref{fig:modelbias} is a form of
\emph{regression towards the mean}, and it is present even in a simple univariate
regression with a large amount of noise \cite{Chernick2003}.  The slope in Figure
\ref{fig:modelbias}, which plots true fitness $y$ against the residual
$\hat{y}-y$, arises because the quadratic statistical model cannot estimate the
higher-order (cubic) interactions, which effectively act as noise in the
regression.  In fact, the slope in the figure equals $1-R^2$, where $R^2$ denotes
the coefficient of determination of the original regression (see Material and
Methods for a derivation).  Whether regression towards the mean is viewed as bias
depends on the interpretation of the statistical model. If one assumes that the
model cannot be improved by adding any more predictor variables, i.e.~that the
noise is caused by purely random factors, as opposed to unknown systematic
factors, then the regression results are unbiased and the observed negative slope
between $\hat{y}-y$ and $y$ simply reflects the fact that the regression cannot
estimate the noise.  However, in situations when there is a systematic signal that
is missing from the statistical model, such as when fitting a quadratic model to a
fitness landscape that contains higher-order interactions, then the regression is
biased towards the mean in proportion to the amount of variance that is not
explained by the model. This phenomenon is not caused by measurement noise but by
the omission of relevant variables.  In the experiments summarized in
Fig.~\ref{fig:modelbias} there is no measurement noise in the fitnesses, and so
the negative slope shown in Figure \ref{fig:modelbias} reflects a true bias: the
mis-specified model over-estimates low fitnesses and under-estimates high
fitnesses.

With carefully tuned parameters, other forms of mis-specified model bias are also
possible, see Fig.~\ref{fig:modelbias2}.  In any case, whatever form it takes,
mis-specified model bias has consequences for how accurately
the landscape inferred from an experiment will reflect the amount of epistasis in
the true landscape or predict the dynamics of adaptation, as we will demonstrate
below.



\subsection{Bias arising from Wright-Fisher sampling}

A third difficulty that arises when fitting a statistical model to measured
fitnesses is the presence of correlations between observed states of loci in
sampled genotypes. An adapting population does not explore sequence
space randomly, but rather is guided by selection towards higher-fitness genotypes.
Sequences sampled from a population under selection will thus tend to have
correlated loci, due either to shared ancestry or due to epistasis. 

Correlated variables do not themselves bias inferred coefficients (at least,
when the model is specified correctly), but they can inflate the variance of those
estimates \cite{Kutner2003}. Predictions from the inferred model are not affected,
in expectation, provided the new data have the same correlations as in the
original training data.  However, in the context of the expansion in Eq.
\ref{eq:quad}, if there are correlations between the included variables, then there are
also correlations between the omitted higher-order interactions and the included
variables. Thus, sampling from a Wright-Fisher (WF) population will exacerbate
the mis-specified model bias.  As a result, the inferred $\beta_i$'s and
$\beta_{ij}$'s will be further biased and so too will the inferred fitnesses.

\begin{figure}
\begin{centering}
\includegraphics[width=1\columnwidth]{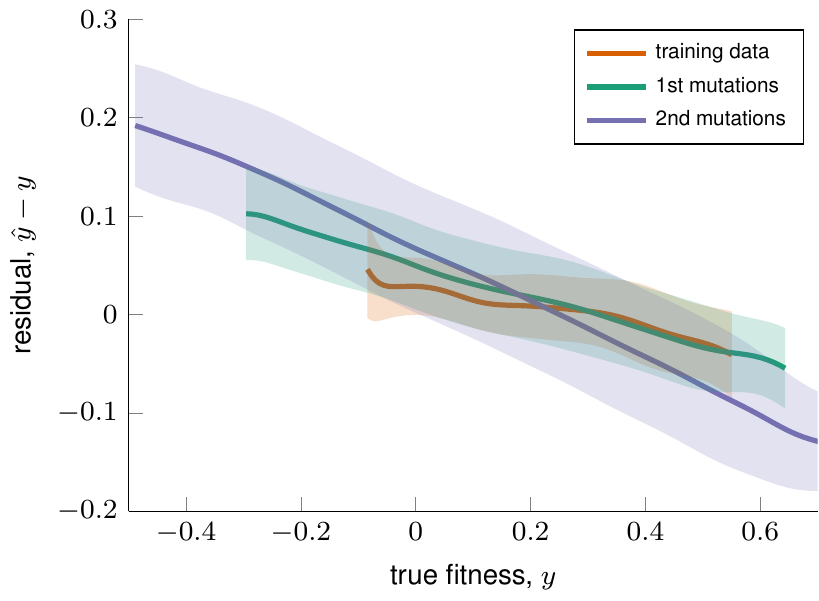}
\par\end{centering}
\caption{Fitness-dependent bias due to both model mis-specification and penalized
regression for genotypes sampled from a Wright-Fisher population under selection.
The predicted fitnesses ($\hat{y}$) were computed from cross-validated training data (red),
for genotypes sampled one mutation away from the training data (purple), and for genotypes sampled two
mutations away from the training data (cyan). The true fitnesses ($y$) are
determined by a cubic polynomial 
fitness landscape on $L=20$ sites with $v_1=0.6$, $v_2=0.3$, $v_3=0.1$, and $\sigma^2_y=0.05$. 
Genotypes for fitting the quadratic statistical model were sampled from the
population after 100 generations of Wright-Fisher evolution, 
with mutation rate $U=10^{-3}$ and population size $N=10^{6}$ (see Methods).  
\label{fig:wfsample2}}
\end{figure}

To illustrate biases that arise from sampling a population under
selection, we simulated Wright-Fisher populations for 100 generations on
cubic polynomial and $NK$ landscapes.  We used large population sizes and mutation
rates to produce a large amount of standing genetic diversity for sampling
genotypes (see Methods). The resulting fits exhibit very high $R^2$
values, in many cases even larger than fits to randomly sampled
genotypes.  But the large $R^2$ values and apparent lack of bias in
the training data are very misleading when the model is misspecified, i.e.~
when the true landscape contains higher-order interactions. When predicting
fitnesses of genotypes just one or two mutations away from the training data we
find again large biases and large variance (polynomial landscape Fig.~\ref{fig:wfsample2}
and $NK$ landscape Fig.~\ref{fig:wfsample_NK}).  As before,
the resulting bias tends towards intermediate fitness values.

\subsection{Extrapolative power to predict fitnesses}

One of the motivations for fitting a statistical model of a fitness landscape
is to predict the fitnesses of genotypes that were not sampled or assayed in the
original experiment. This immediately raises the question, how much predictive
power do such statistical fits have, and how does their power depend upon the form
of the underlying landscape from which genotypes have been sampled, as well as the
form of the fitting procedure?

Although extrapolation is easy to visualize in a linear regression with one
component of $x$, it cannot be plotted as easily in high dimensions, where 
it is sometimes called \emph{hidden extrapolation} \cite{Kutner2003}. In the discrete,
high-dimensional space of genotypes, no genotype is between any two other
genotypes, so that every prediction is in some sense an extrapolation rather than
an interpolation. 

\begin{figure}
\begin{centering}
\includegraphics[width=1\columnwidth]{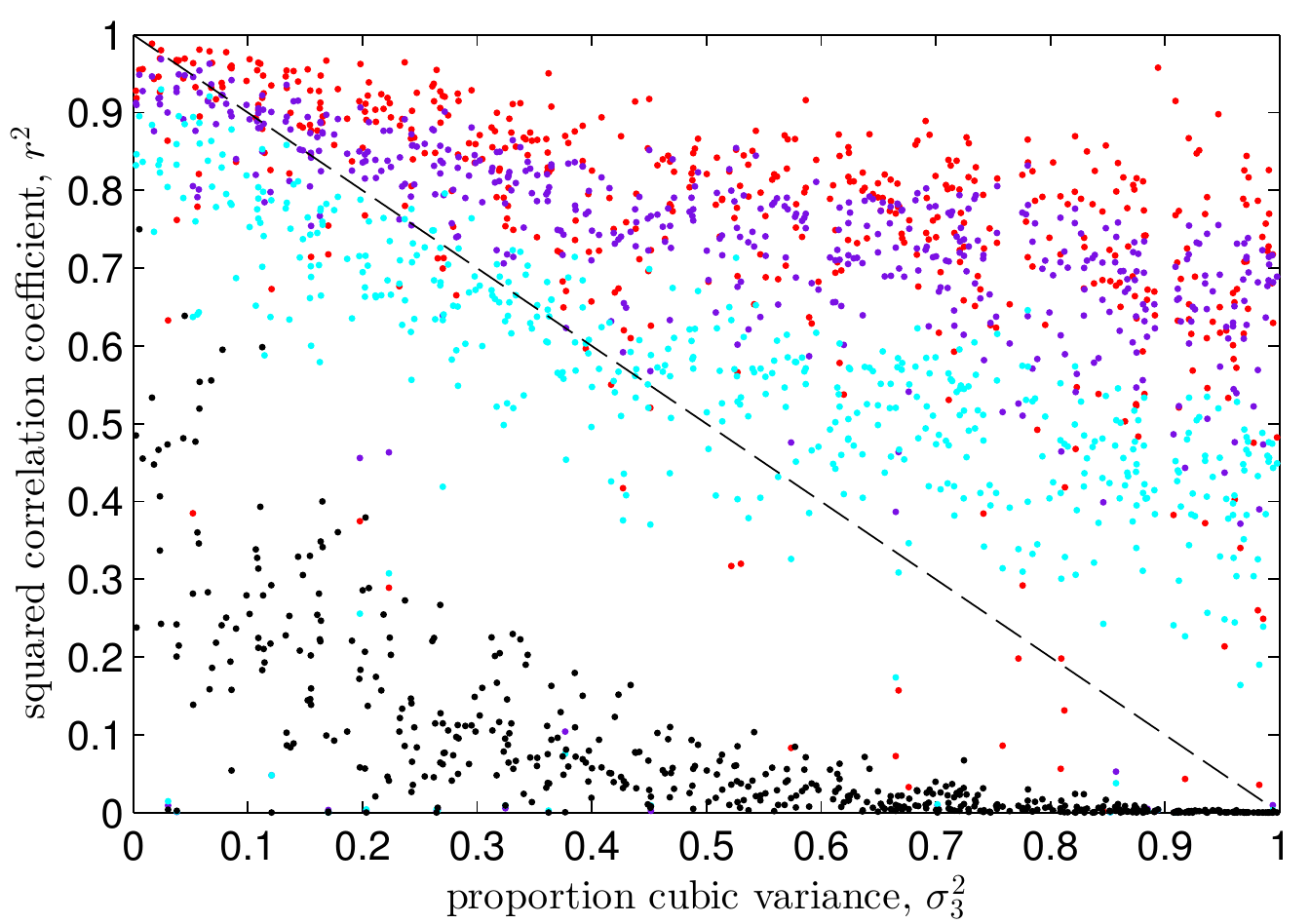}
\par\end{centering}
\caption{Predictive power as measured by squared correlation coefficients between
true and inferred fitnesses for 500
regressions trained on samples from a Wright-Fisher population under
selection.  Cross-validated squared correlation coefficients from the training
data (red) indicate that the fit
obtained from sampling a population under selection can be more accurate than
expected from a regression on randomly sampled
genotypes (dashed line). Predictive power for fitnesses of unsampled genotypes 
are quantified by the squared correlation coefficients between model predictions
($\hat{y}$) and true fitnesses ($y$), 
for sequences that are one mutation away from the training data (purple), two
mutations away from the
training data (cyan), and for random sequences (black). Genotypes used as training
data were sampled from a WF
population after 100 generations of evolution with mutation rates $U=10^{-3}$ and
population size $N=10^{6}$ (see Methods).
Landscapes were each instances of a cubic polynomial form, with $v_3$ values ranging from zero
to one ($x$-axis), with $v_2$ drawn uniformly in range $\{0,1-v_3\}$, and
with $v_1=1-v_3-v_2$. The number of unique sequences sampled from each WF population
varied from 34 to 603 (not shown).
\label{fig:wfsample3}}
\end{figure}

Given experimental data, it may be hard to determine if a model is extrapolating
accurately or not \cite{Otwinowski2013}. Here, we 
quantify the accuracy of extrapolation explicitly using mathematical fitness landscapes.
Fig.~\ref{fig:wfsample3} illustrates the ability of statistical fits to predict
fitnesses of genotypes that were not sampled in the training data, for a range of
models and for regressions with varying degrees of mis-specification. This figure
quantifies the amount of error when predicting the fitnesses of genotypes that are
one or two mutations from the training data, as well as for predicting fitness of
random, unsampled genotypes. Away from the training data, the bias and variance
increase with each mutation, as reflected by the lower squared correlation
coefficients between true and inferred values. The predictions are progressively worse as the amount of model
mis-specification increases. 

A statistical model that has a good fit to the training data, i.e.~a high $R^2$,
does not necessarily imply that the model can make accurate predictions, especially if
there is over-fitting. In fact, Fig.~\ref{fig:wfsample3} shows that
even a high \emph{cross-validated} $R^2$ can be
misleading in the context of predicting unobserved fitnesses when the model is mis-specified.

It is interesting to compare the extrapolative power of landscapes fitted to genotypes
sampled from a Wright-Fisher population, versus genotypes sampled randomly.
The dashed line in Fig.~\ref{fig:wfsample3} indicates the
expected $R^2$ for regressions fitted to randomly sampled genotypes. On the one hand,
predictions that are local, i.e.~within a few mutations from the training data,
typically have a higher $R^2$ for a model trained on WF-sampled genotypes compared to
a model trained on random genotypes. On the other hand, predictions that are far
from the training data (i.e.~predictions for random genotypes),
typically have much lower $R^2$ for a model trained on WF-sampled genotypes
compared to a model trained on random genotypes. Thus, samples from a
Wright-Fisher population produce a more biased model, even of the training data,
but may nonetheless produce better predictions for local unsampled genotypes, compared to
a model fitted to random genotypes \cite{Otwinowski2013}.


\subsection{Biases influence the amount of epistasis in the inferred landscape}

The dynamics of an adapting population depend fundamentally on the form of
epistasis, that is, the way in which fitness contributions from one locus depend
upon the status of other loci.  Indeed, one of the primary goals in fitting a
fitness landscape to empirical data is to quantify the amount and form of its
epistasis, in order to understand how adaptation will proceed.

Given the two sources of biases discussed above, which are inherent to fitting
fitness landscapes to empirical data and exacerbated by sampling from populations
under selection, the question arises: how do these inferential biases influence
the apparent form of epistasis in the fitted landscape? In this section we address
this question by comparing the form of epistasis in the true, underlying fitness
landscape to the form of epistasis in the inferred landscape obtained from fitting
a quadratic model to sampled genotypes.  There are several measures of epistasis
known to influence the dynamics of adaptation. We will focus on three measures
commonly used in the experimental literature on epistasis.

One measure of epistasis, which reflects the degree of predictability in
adaptation, is to reconstruct all the possible genetic paths between a low-fitness
ancestral genotype and a high-fitness derived genotype sampled from an
experimental population
\cite{Weinreich2005,Weinreich2006,Poelwijk2007,O'Maille2008,Lozovsky2009,Tracewell2009,Kogenaru2009,Novais2010,Chou2011}
The proportion of such paths that are "accessible", or monotonically increasing in
fitness, is then a natural measure of epistasis. When this proportion is high, 
many possible routes of adaptation are allowable, suggesting that the
evolutionary trajectory cannot be easily predicted in advance. Whereas when
this proportion is small, it suggests that the evolutionary trajectory is more
predictable, at least in a large population.

\begin{figure}
\begin{centering}
\includegraphics[width=1\columnwidth]{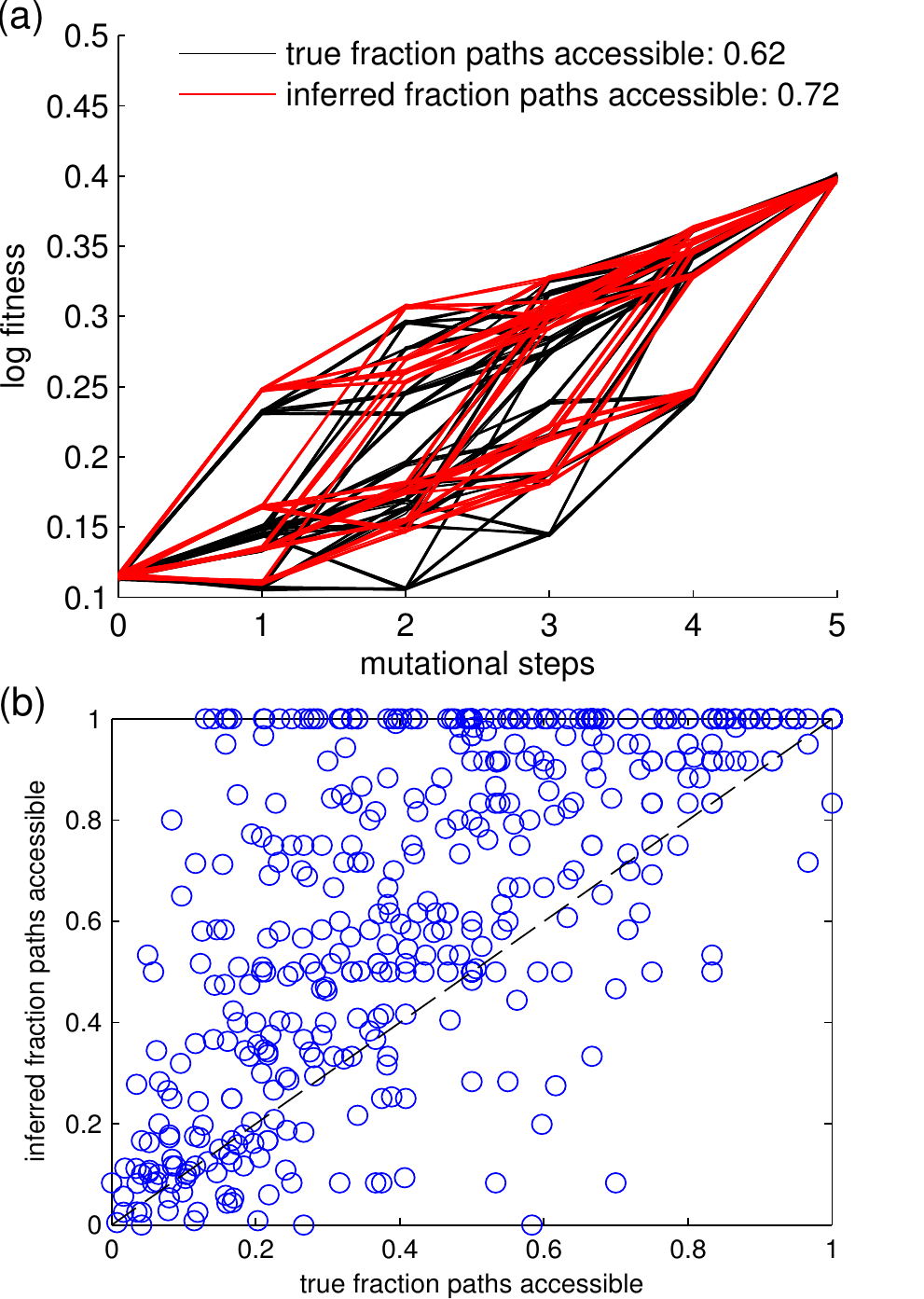}
\par\end{centering}

\caption{Bias in the inferred fitness landscape results in bias in a standard
measure of epistasis: the proportion of "accessible paths", paths that are
monotonically increasing in fitness between an ancestral and an adapted genotype.
(a) All possible mutational paths between a low- and a high-fitness genotype
separated by 5 mutations, under the true (black) and inferred (red) fitness
landscape. The bias towards the mean in the inferred landscape tends to  reduce
the number of fitness valleys, and thereby increases the number of paths
``accessible" to evolution.
(b) This bias in the apparent proportion of accessible paths occurs generally,
across many independent draws of the true underlying fitness landscape.  For each
landscape, we simulated populations that began monomorphic for a low-fitness
genotype and then evolved for 100 generations under selection (see Methods). The
resulting most-frequent genotype was used as the derived genotype, and the final
population was used to fit a quadratic model of the fitness landscape.  All
mutational paths between ancestral and derived genotypes were evaluated, provided
the two genotypes differed by 5, 6, or 7 substitutions.  The graph compares the
fraction of accessible paths for the true ($x$-axis) and inferred ($y$-axis)
landscapes.
The inferred landscapes tend to over-estimate the proportion of accessible paths,
compared to the true landscape: the proportion of accessible paths was
overestimated 3.9-times more often than it was underestimated.  In all cases the
true landscape was cubic polynomial with $v_1=1/3$, $v_2=1/6$, $v_3=1/2$, and
$\sigma^2_y=0.01$. WF simulation parameters: $U=10^{-3}$, $N=10^6$, with 500
generated landscapes and simulations.  \label{fig:paths2}}
\end{figure}

To generate data similar to what would arise in an evolution experiment, we ran
Wright-Fisher simulations on mathematical fitness landscapes (see Methods).  Each
population began monomorphic and the most populated genotype at the end of the
simulation was taken as the derived genotype, which typically contained between 5
to 7 mutations compared to the ancestral genotype.  Genotypes and their associated
fitnesses were sampled from the population after 100 generations and used to fit a
quadratic model of the landscape.  Figure \ref{fig:paths2}A shows an example of
all the mutational paths between the ancestral and derived genotypes separated by
5 mutations, for both the true and the inferred fitnesses.  Since low fitnesses
are likely to be overestimated, and high fitnesses are likely to be
underestimated, the bias in the inferred landscape tends to eliminate fitness
valleys. As a result, the number of accessible paths is higher in the inferred
landscape than it is in the true underlying landscape (Fig.~\ref{fig:paths2}a).
Epistasis appears to be less severe, and adaptation appears able to take more
paths, than it actually is. This effect occurs systematically, as we have observed
it over many realizations of different underlying fitness landscapes
(Fig.~\ref{fig:paths2}b).

We also investigated two other measures of epistasis: the number of local maxima
in the fitness landscape
\cite{Kauffman1987,Kauffman1989,Lobkovsky2011,Szendro2013,Jimenez2013}, and the
prevalence of sign epistasis between pairs of mutations
\cite{Poelwijk2007,Kvitek2011a,Szendro2013,Maharjan2013}. Both of these quantities
are global measures of epistasis, which depend upon the entire landscape,
as opposed to the local measure of accessible paths between an ancestral and
a derived genotype. Generally speaking,
local maxima tend to slow adaptation towards very high fitnesses, even though
valley-crossing can occur in large populations
\cite{Iwasa2004,Weinreich2005a,Weissman2009,Covert2013}. Sign epistasis occurs
when the fitness effect of a mutation at one site changes sign depending upon the
status of a second site. Reciprocal sign epistasis is a subset of sign epistasis,
and it occurs when the second site also has sign epistasis on the first site. 

Fig.~\ref{fig:max_signep} compares the true and inferred amounts of these two
global measures of epistasis. Both quantities can be either underestimated or
overestimated by the inferred landscape, depending on the circumstances.  When the
model is miss-specified (Fig.~\ref{fig:max_signep} bottom row), and without
penalization or local sampling of genotypes, both measures of epistasis are
heavily underestimated, since the model is unable to capture the local maxima and
sign epistasis caused by three-way interactions.  But when genotypes are sampled
locally (Fig.~\ref{fig:max_signep} top row), i.e.~sampled within a few mutations
around a focal genotype, and penalized regression is applied, then the inferred
landscape is influenced both by penalization bias and extrapolation error.  The
penalization bias tends to smooth the inferred landscape and eliminate local
maxima; wherease extrapolation adds noise to the estimated fitnesses and it may
create spurious local maxima. Which of these two effects dominates depends on the
amount of data sampled and how it is distributed.  Sign epistasis appears to be
less sensitive to extrapolation and bias when the model is well-specified,
presumably because it depends only on the signs of effects and not magnitudes. In
all cases considered, however, the inferred landscapes exhibit systematically
biased global measures of epistasis.

\subsection{More realistic landscapes}

\begin{figure}
\begin{centering}
\includegraphics[width=1\columnwidth]{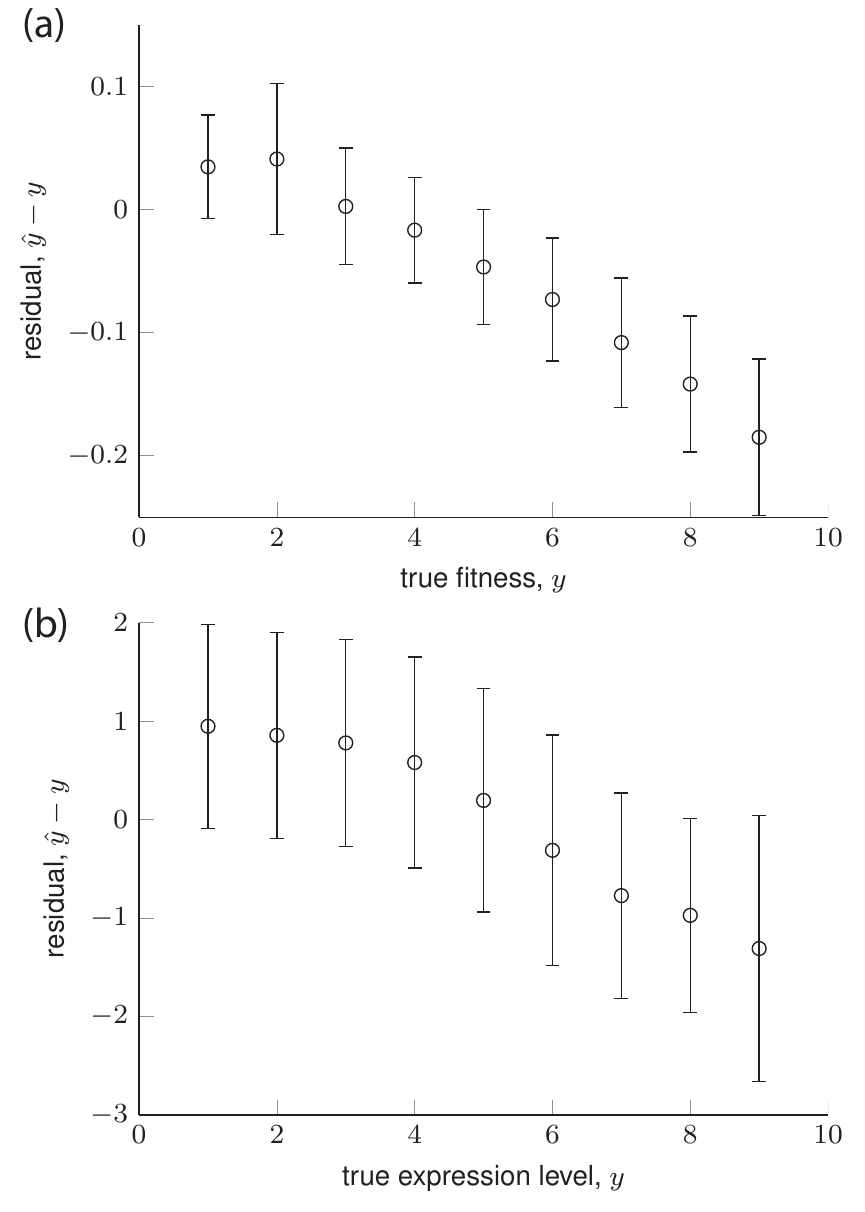}
\end{centering}
\caption{\label{fig:RNAKinney} Bias arising in a quadratic fit to (a)  computational
RNA-folding landscape (see Methods), and (b) regulatory sequence binding landscape from Kinney et al.
\cite{Kinney2010a,Otwinowski2013} (see Methods). The data are discrete in true fitnesses, $y$. Circles
indicate means of distributions of residuals within each bin, and error bars indicate
standard deviations. The quadratic fit exhibits the same type of bias towards the
mean fitness as observed in $NK$ and polynomial fitness landscapes.}
\end{figure}

To complement the studies above, which are based on mathematically constructed
fitness landscapes, we also investigated more realistic landscapes: computational
RNA-folding landscapes and empirical data relating regulatory sequences and
expression levels (i.e.~a regulatory sequence binding landscape, see Methods).
The RNA-folding landscape  and the regulatory sequence binding
landscape (Fig.~\ref{fig:RNAKinney}) both exhibit the same form of bias that we observed
in the mathematical fitness landscapes. 

In the case of the RNA-folding landscape (see Fig.~\ref{fig:RNAKinney}a) there is no
measurement error and sufficient data to avoid the need for penalized regression.
Thus, the bias towards the mean fitness seen in Fig.~\ref{fig:RNAKinney}a with $R^2=0.32$
is due entirely to model misspecification: the quadratic model does not capture
some higher-order interactions that influence RNA folding. The regulatory sequence
expression level data (Fig.~\ref{fig:RNAKinney}b), on the other hand, contain some measurement noise which
comprises about 10-24\% of the variance \cite{Otwinowski2013}, whereas the $R^2$
for statistical model is $0.57$. These numbers suggest that higher-order
interactions bias the predictions made from the statistical model of the
regulatory sequence binding landscape, as well, at least to some extent. 

The form of the biases observed when fitting a quadratic model to these realistic
fitness landscapes (Fig.~\ref{fig:RNAKinney}) are similar to the biases
observed when fitting quadratic models to $NK$ landscapes or to polynomial
landscapes. Therefore we expect that these biases will have similar consequences
for measures of epistasis.

\subsection{Reducing bias when fitting fitness landscapes}

\begin{figure}
\begin{centering}
\includegraphics[width=1\columnwidth]{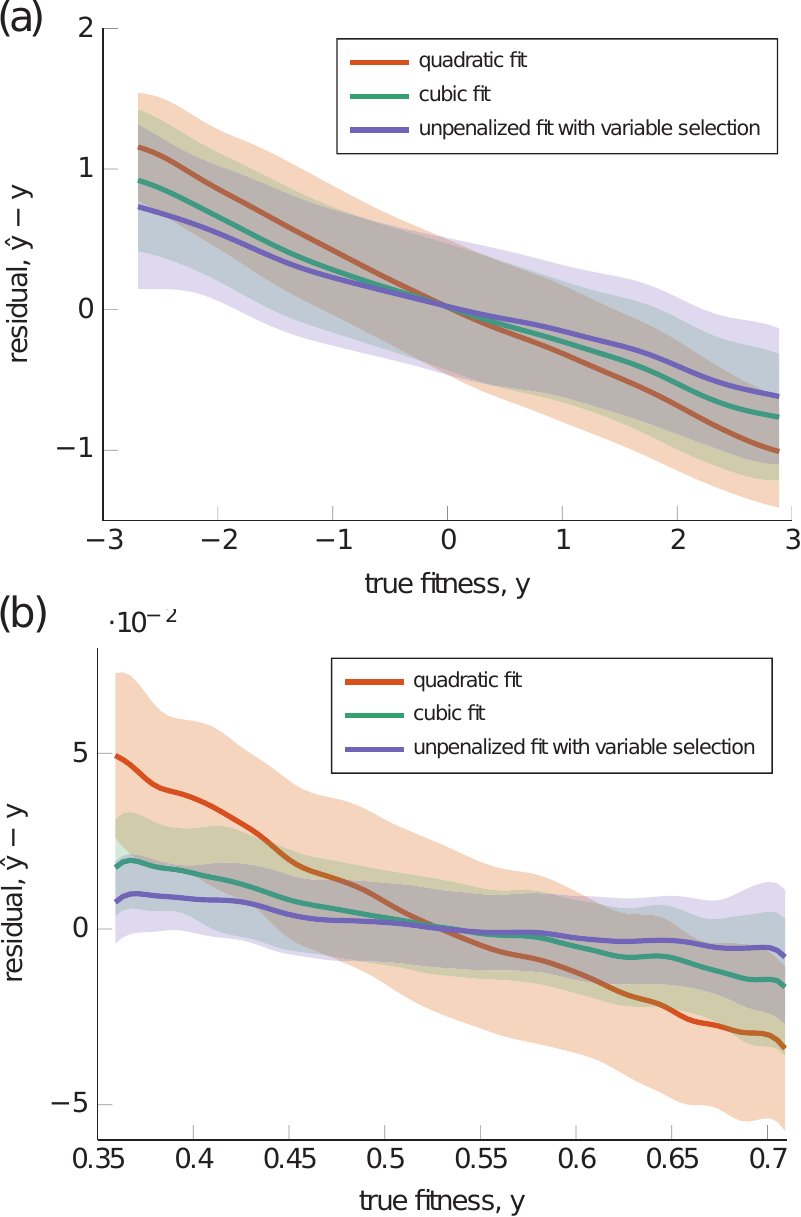}

\par\end{centering}
\caption{\label{fig:unbias} Reducing bias when fitting a statistical model to
(a) cubic polynomial landscape, and (b) $NK$ landscape.  Bias in the
inferred fitnesses can be reduced by adding third-order interactions to the
statistical model (quadratic fit in red, compared to cubic fit in green).
Although the green model is correctly specified, some bias still remains due to
the penalized regression  ($p=1,351$ parameters fit with $N_{train}=800$ data
points). Further reduction of bias
can be achieved by selecting model variables with LASSO \cite{Tibshirani1996,Friedman2001}, and
then performing an unpenalized regression only the selected variables (blue). The variable
selection step may omit some important variables, especially when the true
landscapes includes a large number of higher-order interactions, as in the cubic
polynomial case (a). If the true landscape has a sparse set of interactions (e.g.~the
$NK$ landscape, panel b), then bias can be removed almost entirely by this two-step
procudure.  
(a): cubic polynomial landscape with
$v_1=1/3$, $v_2=1/6$, $v_3=1/2$. (b): $NK$ landscape with $K=2$.
A test set of 5,000 random genotypes was used to compare the predicted ($\hat{y}$) and
true ($y$) fitnesses.
} \end{figure}

Bias arising from a mis-specified model may be reduced by adding relevant
missing variables. In the context of the expansion in Eq.~\ref{eq:quad} this
requires adding higher-order interactions such as triplets of sites,
quartets, etc. In practice, this approach is often infeasible because the
number of such interactions is extremely large and relatively few of them may be
present in the data.  In fig.~\ref{fig:unbias}a we show that the bias in the
statistical model can be reduced by fitting a model that includes three-way
interactions.  However, there is still some residual bias, due to penalized
regression. Thus, incorporating additional predictor variables effectively trades bias due
to model mis-specification for bias due to a more severe penalized regression.

It is not always necessary to use penalized regression, especially when
higher-order interactions are sparse. In such cases it may be appropriate first to
select a limited number of relevant variables, and then to use standard regression
that avoids the bias associated with penalization.  Selecting the variables to use
in the statistical model may be done "by hand" based on prior knowledge, or by
statistical methods such as LASSO \cite{Tibshirani1996,Friedman2001} (see
Methods). This approach is expected to perform well only when a relatively small
number of variables/interactions are in fact present in the true fitness
landscape.

As a proof of principle, we have shown how LASSO followed by standard unpenalized
regression reduces bias in fits to the cubic polynomial and $NK$ landscapes.
Notably, LASSO is a form of penalized regression that favors sparse solutions.
We do not use LASSO in this procedure to make predictions, but rather to select
the  variables to retain for the eventual unpenalized regression. 
The cubic polynomial landscape contains interactions between
all triplets of sites, and so the variables retained by LASSO are expected to omit some
important variables, leaving some bias. By contrast, the $NK$ landscape has sparse
interactions and the resulting fit of this two-step procedure is therefore
expected to be far less biased. Both of these expectations are confirmed in
Fig.~\ref{fig:unbias}. Although this approach may have the benefit of reducing
bias in the inferred fitnesses, it will not improve the overall $R^2$ of the fit,
or the extrapolative power.

\section{Discussion}

Our ability to measure the genotype-fitness relationship directly in experimental
populations is advancing at a dramatic pace. And yet we can never hope to measure
but a tiny fraction of an entire fitness landscape, even for individual proteins.
As a result, there is increasing need to fit statistical models of landscapes to
sampled data. Here we have shown that such statistical fits can warp our view of
epistasis in the landscape and, in turn, our expectations for the dynamics of an
evolving population.

We have identified two distinct sources of biases: penalized regression and model
mis-specification. Our analysis of the effects of penalized regression have been
performed using ridge regression, but the same qualitative results hold for LASSO
regressions (Figs. \ref{fig:regbias_poly_lasso}, \ref{fig:regbias_nk_lasso},
\ref{fig:WF_poly_lasso}, \ref{fig:WF_nk_lasso}, \ref{fig:paths_all_lasso}
analogous to Figs.~\ref{fig:regbias}, \ref{fig:regbias_NK}, \ref{fig:wfsample3},
\ref{fig:wfsample_NK}, \ref{fig:paths2}b), and also, we expect, for other forms of
penalization such as elastic net or the generalized kernel ridge regression in
\cite{Hinkley2011,Kouyos2012}.  Notably, the bias arising from penalized
regression has a slightly different form depending upon whether genotypes are
encoded in the $\{0,1\}$ basis relative to the wild-type, in which case inferences
will be biased towards the wild-type fitness, or the $\{+1,-1\}$ basis, used here,
in which case inferences will be biased towards the average fitness.

Aside from these biases, we have also shown that statistical fits have poor
predictive power.  Even when a fitted landscape exhibits a large cross-validated
$R^2$ value, the fitted landscape generally has poor power to predict the
fitnesses of unsampled genotypes, including genotypes within only a single
mutation of the genotypes used to fit the landscape. 

Common measures of epistasis may be grossly distorted by
statistical fits to sampled data. Interestingly, different measures of epistasis
can be affected differently. For example, the number of paths accessible between
an ancestal and derived genotype will be systematically over estimated in such
fits, suggesting that evolution is less predictable than it actually is.  The
number of local maxima in the inferred landscape can be severly over- or
under-estimated, whereas the prevalence of sign epistasis is less prone to bias
(Fig.~\ref{fig:max_signep}).  Estimates of pairwise sign epistasis may be more
robust because quadratic fits capture pairwise interactions, whereas the number of
local maxima can depend on higher-order interactions. 

The problems of bias towards smoothness due to penalized regression and ruggedness
due to extrapolation error can co-exist in the same dataset, as observed in a
recent study of regulatory sequence binding data \cite{Otwinowski2013}. Which of
these two effects will dominate is unclear, in general, as it depends upon the
underlying landscape and the form of sampling. As a result, it is difficult to
interpret the predictions made by quadratic fits to fitness landscapes, such as 
fits made to HIV data \cite{Kouyos2012}.  Nevertheless, it is often possible to at
least deduce the presence of epistatic interactions from sampled data
\cite{Otwinowski2013, Hinkley2011}.  

In some cases, such as when the true landscape has sparse high-order interactions
between loci, a combination of variable selection followed by unpenalized
regression may ameliorate the biases we have identified. However, the degree to
which this approach will reduce bias will surely depend upon the biological context.
Thus researchers should incorporate as much prior biological knowledge as
possible when choosing a statistical model and fitting procedure. At the very
least, it is important that researchers be aware of the biases inherent in 
fitting statistical
of fitness landscapes to data. 






\begin{materials}

\section{Parameterization of genotypes}
In order to build a statistical model with linear and interacting
terms (eq \ref{eq:quad}), genetic sequences must be encoded as dummy
variables, $x_{i}$. If there is a well-defined wild-type sequence,
then a natural parameterization is using zeros and ones, with the wild-type denoted as all zeros. The coefficients $\beta_i$ are the effect of single mutations, $\beta_{ij}$ are the effects of pairs of mutations, and the constant term $\beta_{0}$ is the inferred wild-type fitness. 
In the case of a population with large diversity and no well-defined reference sequence, the reference-free parameterization
with $\pm1$ may be more appropriate, with $\beta_{0}$ then denoting
the inferred average fitness \cite{Weinberger1991,Neher2011}. In this work we used the  $\pm1$ basis, because a wild-type was not defined, and the polynomial landscapes (see below) are defined in the $\pm1$ basis.

\section{Polynomial landscapes}
We constructed ``polynomial landscapes'' by terminating the expansion (Eq. \ref{eq:quad}) at the third-order and specifying the coefficients in such a way as to control the contributions to the total variation in log fitness that arise from interactions of each order. In particular, the variance of $y$ is
\begin{equation}
\langle y^2 \rangle = \sum_{i}^{L}\beta_{i}^2 + \sum_{i<j}\beta_{ij}^2 + \sum_{i<j<k}\beta_{ijk}^2
\end{equation}
where $\langle . \rangle$ denotes an average over all genotypes in the orthogonal ($\pm1$) basis, and the sums are taken overal all sites, pairs of sites, and triplets. The coefficients $\beta_i$, $\beta_{ij}$, and $\beta_{ijk}$ are chosen from normal distributions with mean zero and variances:
\begin{equation}
\sigma^2_{1}=\sigma^2_y v_1/L,
\end{equation}
\begin{equation}
\sigma^2_{2}=\sigma^2_y v_2/ {L \choose 2},
\end{equation}
\begin{equation}
\sigma^2_{3}=\sigma^2_y v_3/ {L \choose 3},
\end{equation}
where $v_i$ is the fraction of total variance determined by the $i$th order, $L$ is the number of loci, and the total variance is $\langle y^2 \rangle=\sigma^2_y $. For our numerical investigations using the cubic polynomial landscapes we chose $L=20$. $\beta_0$ makes no contribution to the variance or the evolutionary dynamics, and was set to zero.

\section{$NK$ landscapes}
We followed a standard \cite{Kauffman1987,Kauffman1989,Kauffman1993} procedure for constructing $NK$ fitness landscapes. The parameter $N$ denotes
the length of the binary string defining the genotype ($N=L=20$ throughout our analyses) The logarithm of the fitness of a genotype is calculated as
the mean of contributions from each site, which are themselves determined by a table of values each drawn independently from a uniform probability
distribution. When $K=0$, the contribution of a site depends only on its own state: 0 or 1, and not on the state of other sites. When $K>0$, the
contribution of a site depends on its own (binary) allele as well as the states of K other sites, yielding a lookup table with $2^{K+1}$ entries.
Thus, there are in general $N$ lookup tables each with $2^{K+1}$ independently drawn entries, which together determine the contribution of each locus,
based on the status of all other loci. Under such $NK$ models, the fitness effect of a substitution depends strongly and randomly on some fraction of
the genetic background, determined by $K$. $K$ is constant across sites and genotypes for a particular landscape, and the identifies of the $K$ sites
upon which a given locus depends are drawn uniformly from the $N-1$ possibilities. Notably, $K=0$ is an additive landscape, and $K=1$ is additive with sparse pairwise interactions. The amount of total variance in fitness due to the $k$th-order interactions is proportional to $2^{-k}{N \choose k}$ \cite{Neidhart2013a}.

\section{RNA folding landscape}
The RNA-folding landscape was generated by the Vienna RNA software \cite{Lorenz2011a}. 
The target secondary structure was chosen as the most common structure observed in a sample of 10,000 structures 
generated from random genotypes of length 15. The fitness function was defined as $1/(1.05)^{h}$, where $h$ is hamming distance 
to the target in the tree-edit metric. Training data consisted of
$10^4$ random genotypes.

\section{E. coli regulatory sequence binding landscape}
The data consisted of 129,000 sequences of the \textit{E. coli} lac promoter and associated gene expression levels. Each sequence contained 75 nucleotides and contained roughly 7 mutations relative to the wild type. \textit{E. coli} were FACS sorted by expression levels into 9 bins, and the bin numbers serves as the phenotype for fitting the quadratic model. In this case, LASSO was used for regression. For more details see \cite{Otwinowski2013}.

\section{Penalized regression}
A quadratic fit with ridge regression was used (unless otherwise stated), which identifies 
the coefficients that minimize
\begin{equation}
(y-\beta_{0}+\sum_{i}^{L}\beta_{i}x_{i}+\sum_{i<j}\beta_{ij}x_{i}x_{j})^2 + \lambda \sum \beta_{i,ij}^2,
\end{equation}
where the last sum is taken over all coefficients denoted as $\beta_{i,ij}$. The first term is the mean squared error, and the last term is the penalization which biases coefficients towards zero. The free parameter $\lambda$ was determined by choosing the largest $\lambda$ within one standard deviation of the smallest ten-fold cross-validated mean squared error. An alternative form of penalization is LASSO \cite{Tibshirani1996}, which has a penalization term of the form $\sum \vert \beta \vert $. LASSO favors sparse solutions of coefficients, and is useful for picking out important variables. Ridge and LASSO regression were done by Matlab version R2013b.

\section{Wright-Fisher simulations} Monte-Carlo simulations of adaptation were
based on standard Wright-Fisher dynamics \cite{Ewens2004}. A population consists
of $N$ individuals, each with a genotype consisting of a bit string of length 20.
The population replaces itself in discrete generations, such that each individual has
a random number of offspring in proportion to its fitness, which is determined by
its genotype via the fitness landscape. In practice, this is done efficiently with
a multinomial random number generator. Mutations are defined as bit flips, and
they are
introduced in every individual at each generation with probability $U$ per
genome. The number of individuals receiving a mutation is thus binomially distributed,
as double mutations are not allowed. The populations were initialized as
monomorphic for a low-fitness genotype, chosen by generating 100 random genotypes
and picking the one with the lowest fitness. Simulations were run for 100
generations, and the resulting population was reduced to unique genotypes, and
those genotypes, with the corresponding fitnesses, were used as the training data
for regressions. The number of unique genotypes in the population is sensitive to
$U$, $N$, the number of generations, and $\sigma^2_y$. These parameters were
chosen such that there were at least a few hundred unique genotypes, representing
substantial diversity in the population.

\section{Plots of  true fitness versus residuals} The figures plotting true
fitness versus residuals
were produced using
a Gaussian moving window applied to the raw data. For each value of true fitness, $y$, 
a mean and standard deviation was
calculated by weighting all the data points by a Gaussian with a width $\sigma$,
and normalized by the sum of all the weights for the given $y$ value. This procedure
provides a sense of the distribution at a given $y$, without regard to the density of
points. Areas on the
extremes of $y$ had few points to estimate a mean and variance, and they were excluded
if the sum of weights was smaller than the 10\% percentile of the distribution of
all normalization factors. We used the smoothing parameter $\sigma=0.3$ for figures
\ref{fig:regbias}, \ref{fig:modelbias}, \ref{fig:unbias}a, \ref{fig:modelbias2},
and \ref{fig:regbias_poly_lasso}; $\sigma=0.01$ for figures \ref{fig:unbias}b,
\ref{fig:regbias_NK}, \ref{fig:modelbias_NK}, and \ref{fig:regbias_nk_lasso};
$\sigma=0.05$ for figures \ref{fig:wfsample2} and \ref{fig:WF_poly_lasso}; and
$\sigma=0.02$ for figures \ref{fig:wfsample_NK} and \ref{fig:WF_nk_lasso}.

\section{Slope of true fitness versus residuals}
A scatter plot of true fitnesses, $y$, versus estimated fitnesses inferred by regression, 
$\hat{y}$, reflects the quality of the statistical fit. One
can calculate the slope of $y$ versus $\hat{y}$ by using a second regression:
\[
y=\beta\hat{y}+\epsilon
\]
where $\beta$ denotes the slope, found by minimizing the mean squared
error 
\[
\langle\epsilon^{2}\rangle=\langle(y-\beta\hat{y})^{2}\rangle
\]
\[
\frac{d\langle\epsilon^{2}\rangle}{d\beta}=0=\langle2(y-\beta\hat{y})(-\hat{y})\rangle
\]
\[
\beta=\frac{\langle y\hat{y}\rangle}{\langle\hat{y}^{2}\rangle}.
\]
Recall that $y=\hat{y}+r$, where $r$ is the residual from the
initial regression, and $\langle r\hat{y}\rangle=0$ because residuals
are uncorrelated with $\hat{y}$. As a result, we conclude that $\beta=1$, which
simply reflects the properties of
the original linear regression. This result is analogous to plotting residuals on the $y$ axis
and inferred values on the $x$-axis, and observing no relationship.
In the main text, by contrast, we show plots of
the true values $y$ versus the residuals $\hat{y}-y$.
In this case we observe a ``bias'', in that genotypes with large $y$ are
underestimated, and genotypes with small $y$ are overestimated. This type of
bias is a form of regression towards the mean. We can calculate the slope of $y$ versus $\hat{y}$ as follows:
\[
\hat{y}=\beta y+\epsilon,
\]
and with a similar calculation we find \[
\beta=\frac{\langle y\hat{y}\rangle}{\langle y^{2}\rangle}=\frac{\langle y^{2}\rangle-\langle y^{2}\rangle+2\langle
y\hat{y}\rangle-\langle\hat{y}^{2}\rangle}{\langle y^{2}\rangle}=\frac{\langle y^{2}\rangle-\langle r^{2}\rangle}{\langle y^{2}\rangle}.
\]
If we have mean-centered data ($\langle y\rangle=0$), then this slope
equals the coefficient of determination of the initial regression, denoted $R^2$.
Equivalently, the slope in plots of $y$ versus $\hat{y}-y$ equals $1-R^{2}$.

\end{materials}





\begin{acknowledgments}
We thank A. Feder, J. Draghi, D. McCandlish for constructive input; and P. Chesson for clarifying the usage of the word between. J.B.P. acknowledges funding from the Burroughs Wellcome Fund, the David and Lucile Packard Foundation, the U.S. Department of the Interior (D12AP00025), the U.S. Army Research Office (W911NF-12-1-0552), and the Foundational Questions in Evolutionary Biology Fund (RFP-12-16).
\end{acknowledgments}





















\end{article}

\pagebreak
\titlefont Supporting information

\setcounter{page}{1}
\renewcommand{\thepage}{S\arabic{page}}

\setcounter{figure}{0}
\renewcommand{\thefigure}{S\arabic{figure}} 
\pagebreak

\begin{figure}
\begin{centering}
\includegraphics[width=.5\columnwidth]{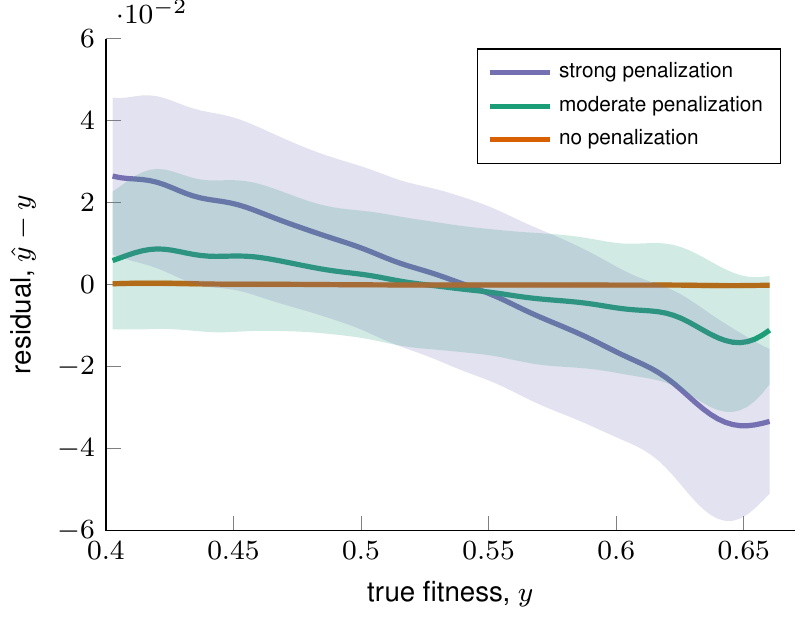}
\par\end{centering}
\caption{
Fitness-dependent bias due to penalized regression. This figure is analogous to figure \ref{fig:regbias}, but with an $NK$ landscape with pairwise interactions ($K=1$ and $N=20$, see Methods). 
\label{fig:regbias_NK} }

\end{figure}

\begin{figure}
\begin{centering}
\includegraphics[width=.5\columnwidth]{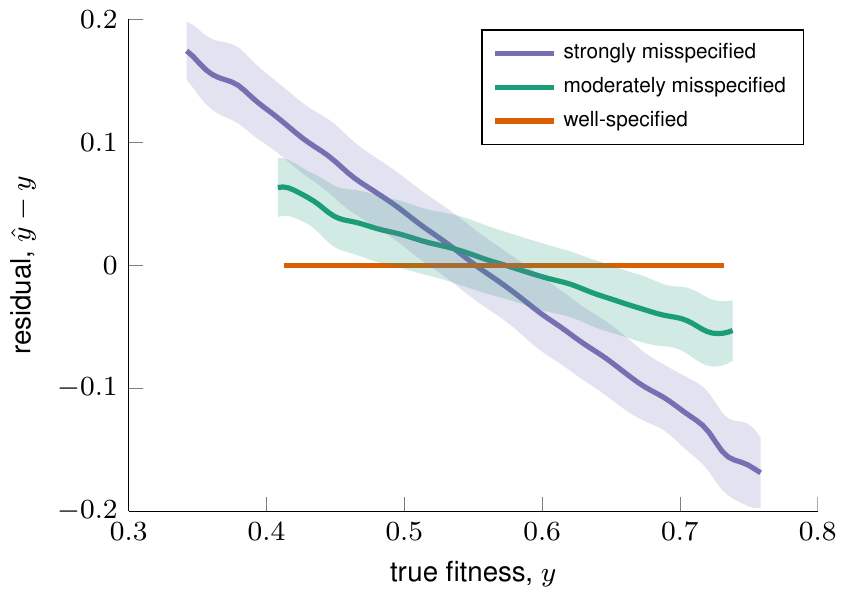}
\par\end{centering}
\caption{
Fitness-dependent bias due to model misspecification. This figure is analogous to figure \ref{fig:modelbias}, but with $NK$ landscapes: $K=1$ (red), $K=3$ (green) and $K=7$ (blue). The larger the value of $K$, the greater the amount of model mis-specification.
\label{fig:modelbias_NK}}
\end{figure}

\begin{figure}
\begin{centering}
\includegraphics[width=.5\columnwidth]{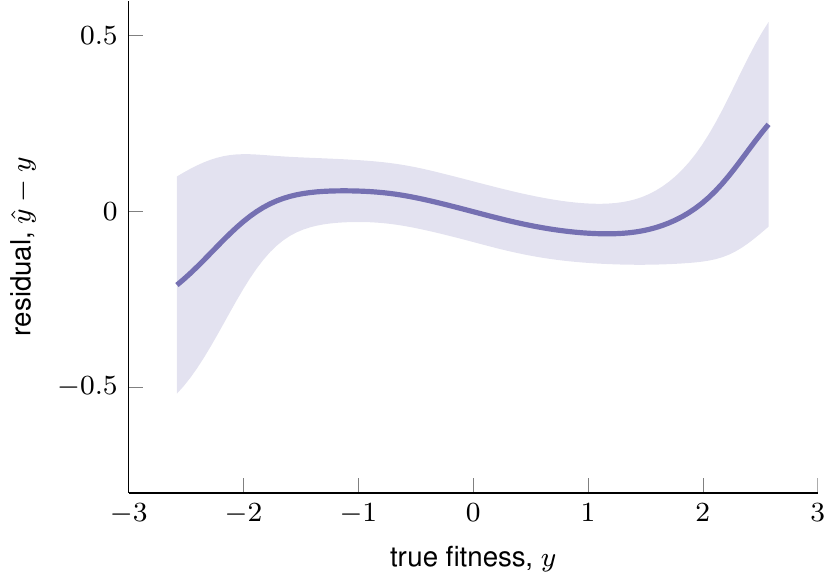}
\par\end{centering}
\caption{\label{fig:modelbias2}
A quadratic model was fit to a cubic polynomial
landscape with large negative additive effects, small mean-zero pairwise effects,
and positive 3-way effects.
The result is a complicated bias in the inferred landscape, with very low fitnesses being
underestimated and very high fitnesses being overestimated. Nevertheless,
the bias from missing, higher-order interactions as in Fig.~\ref{fig:modelbias}
may be a generic feature. When the distribution of interaction effects
is centered around zero, the model tends to be biased towards the
average fitness. }
\end{figure}

\begin{figure}
\begin{centering}
\includegraphics[width=.5\columnwidth]{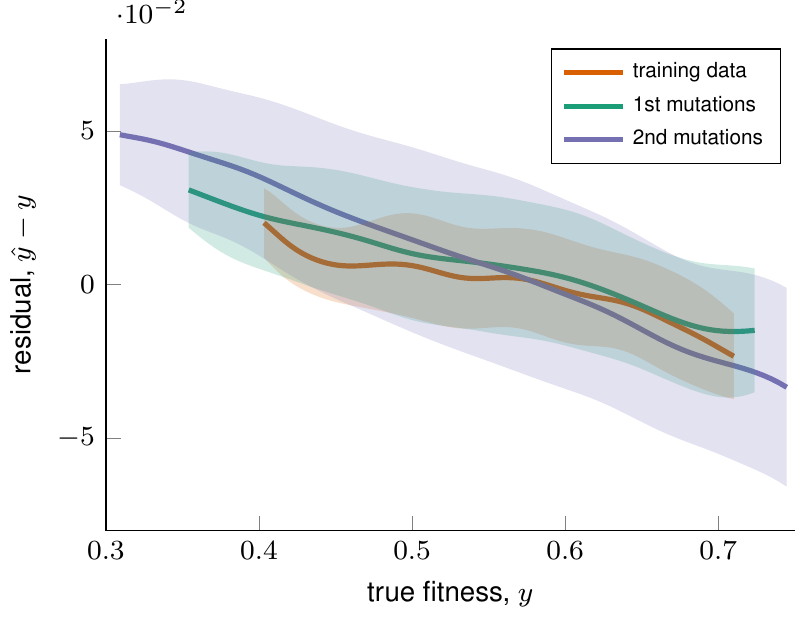}
\par\end{centering}
\caption{ Fitness-dependent bias due to both model misspecification and penalized regression for genotypes sampled from a Wright-Fisher population under selection. This figure is analogous to figure \ref{fig:wfsample2}, but with an $NK$ landscape with $K=3$ and $N=20$.
\label{fig:wfsample_NK}
}
\end{figure}

\begin{figure}
\begin{centering}
\includegraphics[width=.8\columnwidth]{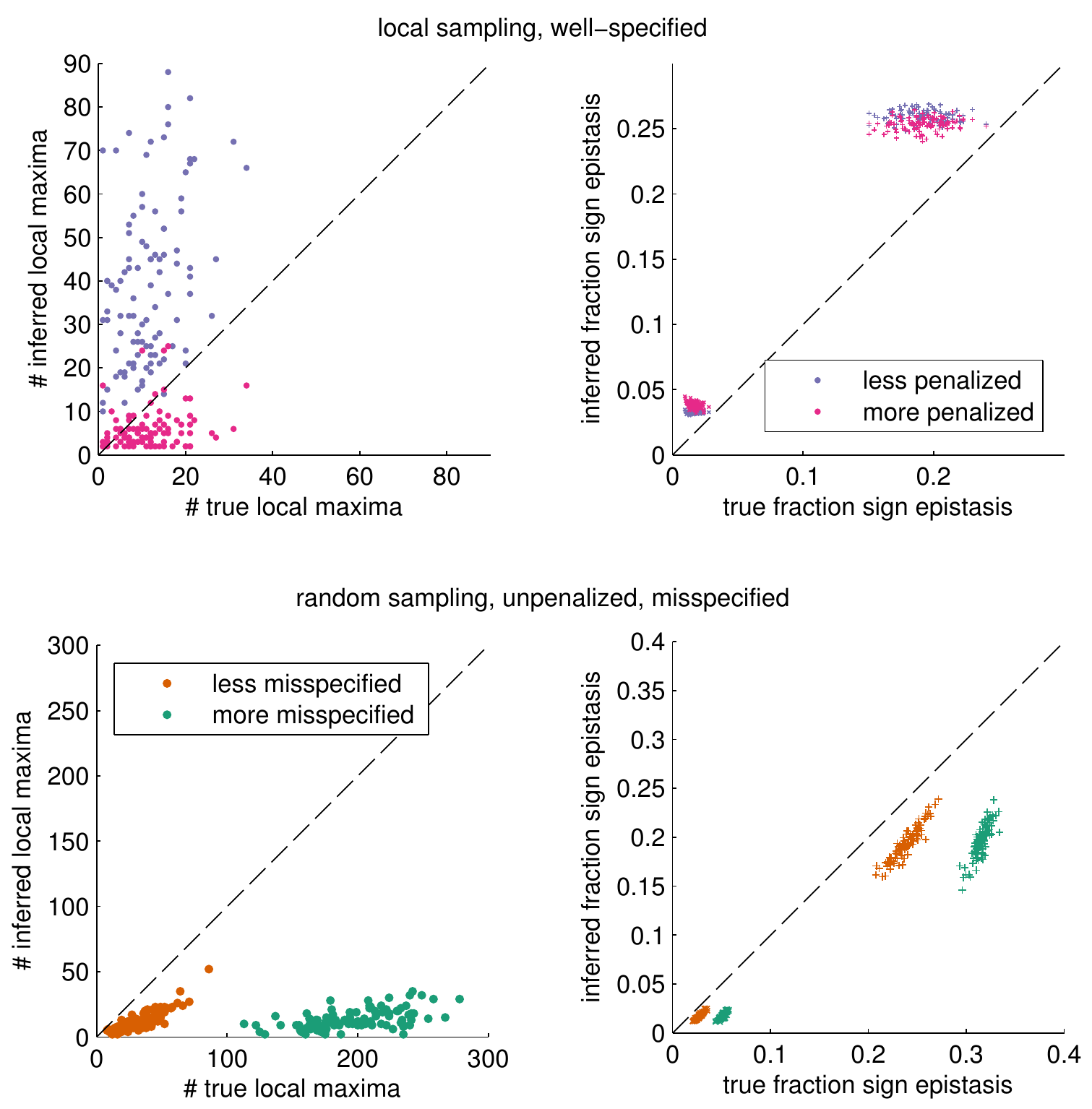}
\par\end{centering}
\caption{\label{fig:max_signep} 
The inferred number of local maxima and the prevalence of sign epistasis may be under- or over-estimated by a statistical model of the fitness landscape fit to sampled data. In the top row the genotypes were sampled ``locally", so that they are close together in hamming distance. The sampled genotypes consist of a focal genotype plus all sequences within one and two mutations, totaling 211 sampled sequences. The true landscape was a quadratic polynomial ($v_1=2/3$, $v_2=1/3$, $v_3=0$), and it was fit with a quadratic model by ridge regression (211 coefficients). This fit did not require a penalized regression, but we imposed some arbitrary values of the penalization parameter, $\lambda$, to demonstrate its effects. When $\lambda=0$, there is no bias (not shown), but when the penalization is modest $\lambda=10$, purple dots), the number of local maxima is over-estimated. The spurious local maxima are likely due to large extrapolation errors, away from the sampled genotypes. When the penalization is yet larger ($\lambda=100$, pink), the penalization bias smooths out the inferred landscape and thus under-estimates the number of local maxima. The fraction of inferred sign epistasis (pluses, upper right cluster) and reciprocal sign epistasis (crosses, cluster in the lower left), is typically overestimated for locally sampled genotypes.  In the bottom row, 5000 genotypes are sampled randomly so that no penalization
is required. However, the true landscape in these cases is cubic polynomial ($v_1=1/3$, $v_2=1/6$, $v_3=1/2$), whereas the statistical model is quadratic, resulting in bias from model mis-specification. In these cases, the number of local maxima and the prevalence of sign epistasis are both under-estimated. In all cases, $\sigma^2_y=1$ and dashed line indicates the line $y=x$.}
\end{figure}

\begin{figure}
\begin{centering}
\includegraphics[width=.5\columnwidth]{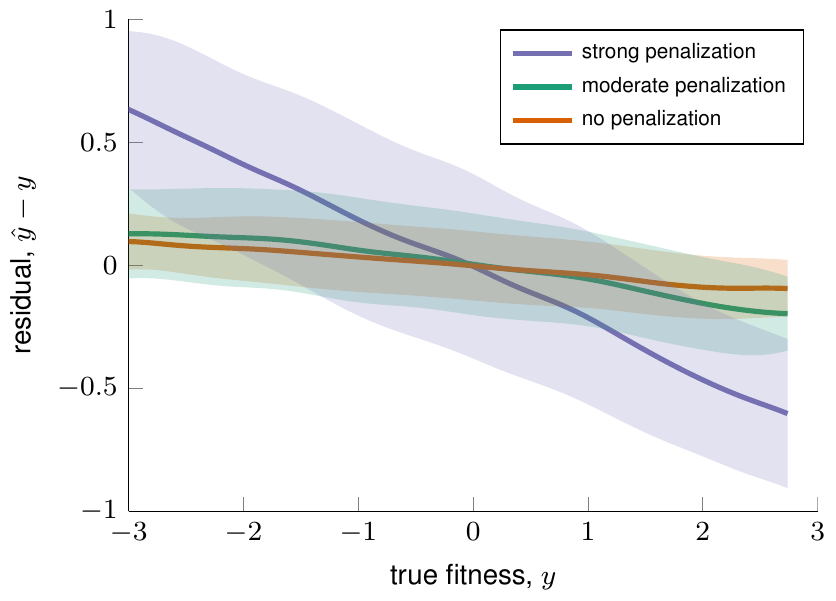}
\par\end{centering}
\caption{ Fitness-dependent bias due to penalized regression in a polynomial landscape. Analogous to figure \ref{fig:regbias}, but with LASSO instead of ridge regression.\label{fig:regbias_poly_lasso}}
\end{figure} 

\begin{figure}
\begin{centering}
\includegraphics[width=.5\columnwidth]{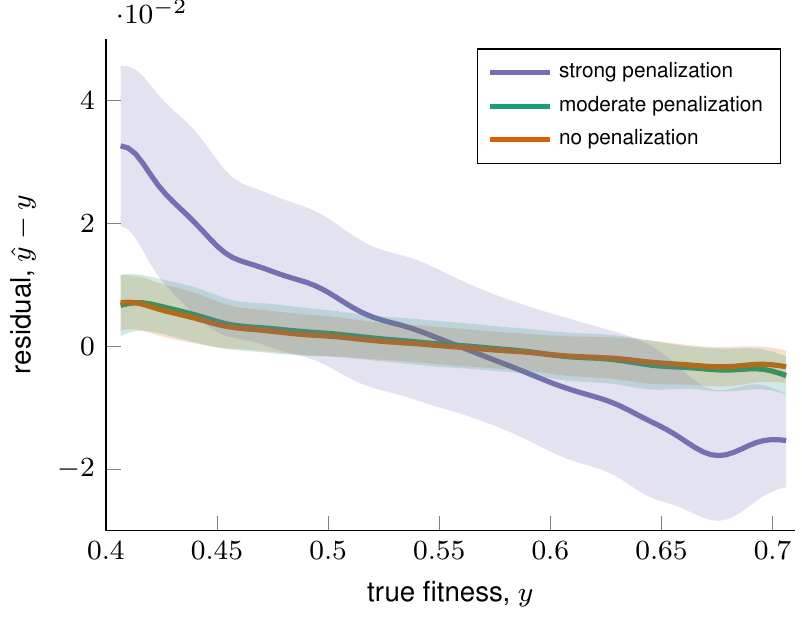}
\par\end{centering}
\caption{ Fitness-dependent bias due to penalized regression in a $NK$ landscape. Analogous to figure \ref{fig:regbias_NK}, but with LASSO instead of ridge regression.\label{fig:regbias_nk_lasso}}
\end{figure} 

\begin{figure}
\begin{centering}
\includegraphics[width=.5\columnwidth]{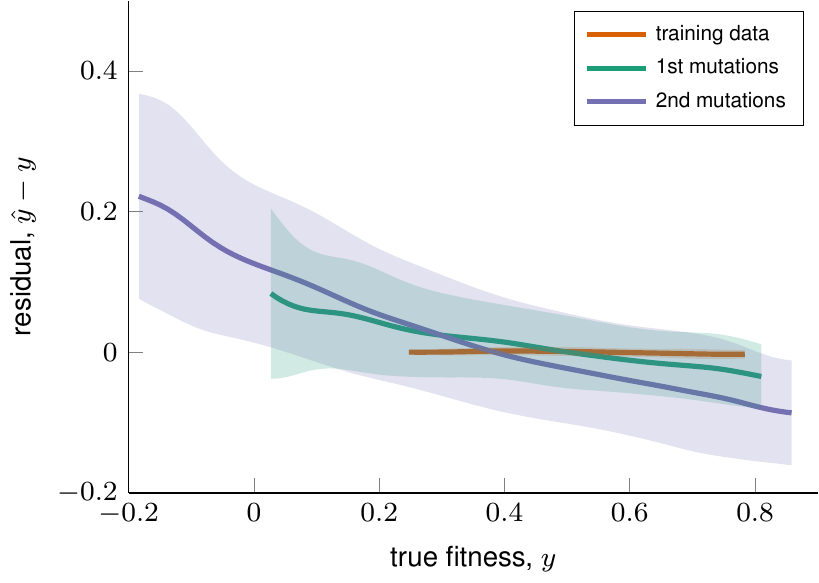}
\caption{ Fitness-dependent bias due to both model misspecification and penalized regression for genotypes sampled from a Wright-Fisher population under selection in a polynomial landscape. Analogous to figure \ref{fig:wfsample2}, but with LASSO instead of ridge regression.\label{fig:WF_poly_lasso}}
\par\end{centering}
\end{figure} 

\begin{figure}
\begin{centering}
\includegraphics[width=.5\columnwidth]{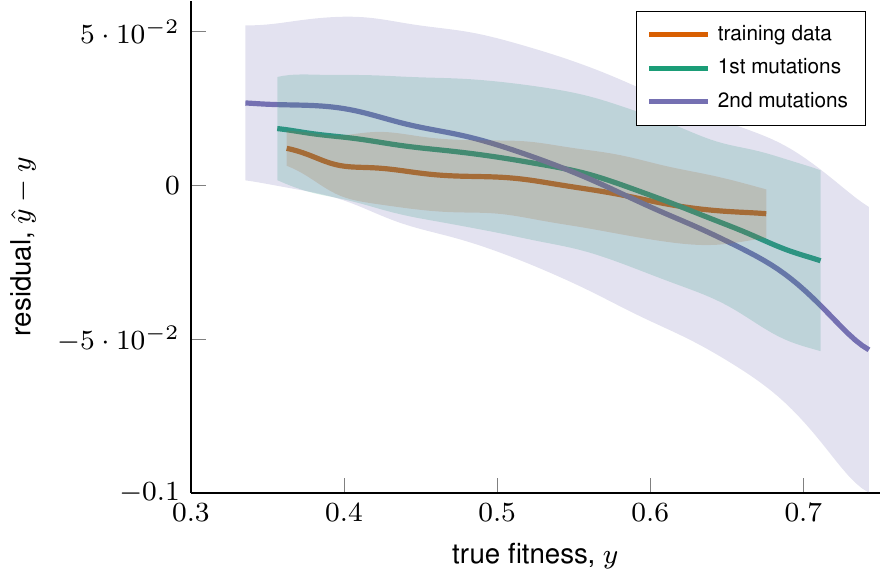}
\caption{ Fitness-dependent bias due to both model misspecification and penalized regression for genotypes sampled from a Wright-Fisher population under selection in $NK$ landscape. Analogous to figure \ref{fig:wfsample_NK}, but with LASSO instead of ridge regression.\label{fig:WF_nk_lasso}}
\par\end{centering}
\end{figure} 

\begin{figure}
\begin{centering}
\includegraphics[width=.5\columnwidth]{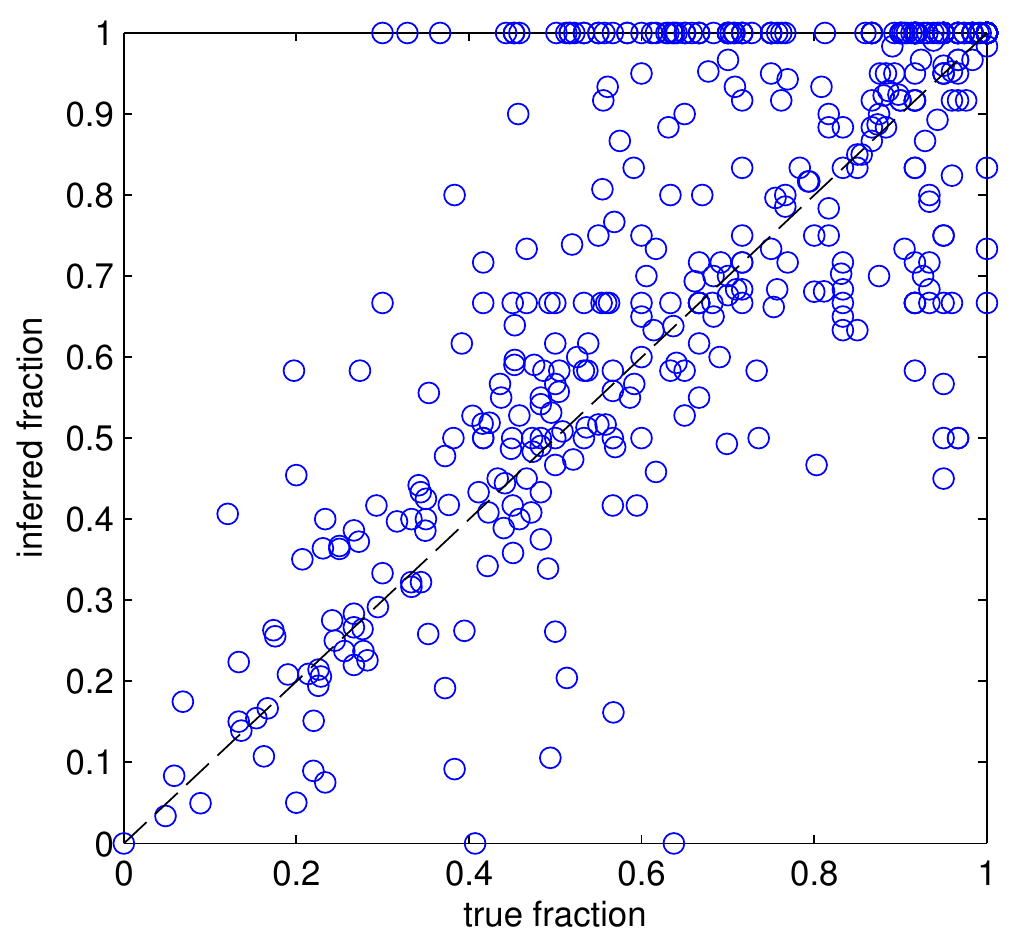}
\par\end{centering}
\caption{ Bias in the proportion of "accessible paths", paths that are monotonically increasing in fitness between an ancestral and adapted genotype. Analogous to figure \ref{fig:paths2}b, but with LASSO instead of ridge regression.\label{fig:paths_all_lasso}}
\end{figure} 
































\end{document}